\documentstyle[prc,preprint,aps,epsf]{revtex}
\begin{document}

\title { Phase-space density in heavy-ion collisions revisited}
\author{Q. H. Zhang$^{1,2}$, J. Barrette$^1$ and C. Gale$^1$}
\address{$^1$Physics Department, McGill university, Montreal QC H3A 2T8, 
Canada\\
$^2$ Computer Science Department, Concordia University, QC H3G 1M8, Montreal,
Canada}
\vfill
\maketitle

\begin{abstract}
We derive the  phase space density of bosons from 
a general boson interferometry formula. 
 We find that the phase space density 
is connected with the two-particles 
 and the single particle density distribution functions. 
If the boson density is large, the two particles 
density distribution function can not be expressed as  
a product  of two single particle density distributions. 
However, if the boson density 
is so small that two particles 
density distribution function can be expressed as a 
product of two single particle density distributions, then 
 Bertsch's formula is recovered.
For a Gaussian model, the effects of multi-particles 
Bose-Einstein correlations on 
the mean phase space density are studied.
\end{abstract}

PACS number(s): 13.60.Le, 13.85.Ni, 25.75.Gz

\section{Introduction}

The principal aim of the study of relativistic heavy-ion collisions 
is the search for evidences of the  
 state of a quark gluon plasma (QGP) in the early stage of 
the reactions\cite{review,rev1}.
A quantity of great interest for the study of the QGP is 
maximum energy density that has been reached in the 
experiment.  This maximum energy density is connected with 
the final state energy density and phase space volume.
For ultra high 
energy collisions, more than eighty percent of final state particles are pions; 
therefore, it is 
very important to measure  the phase space density 
of pions which can be used to infer the 
energy density in the early stage of the collisions\cite{Zhang01}.
Besides this, it is very important to estimate the phase space density 
 also for the following reasons: 
(1) If the boson phase space density is very large, then 
pions will tend to stay in the same state and pion 
condensate may occur.  (2): if the density of pions is very large, 
then the mean free-length among pions will be small; 
therefore, the evolution of pions  
in the final state should be described by the 
hydrodynamical equation\cite{Shur}.

Bertsch suggested a method which uses pion interferometry 
measurements\cite{HBT} to calculate the   
mean pion phase space density several years ago\cite{Bertsch}. Since then 
several calculations have been done at AGS energy\cite{Ba97} and 
SPS energy\cite{Bertsch,HC99,WX,TWH99,Murry,TH99}. 
It has been found that the phase space density of 
pions is very low at AGS and 
SPS energies\cite{Bertsch,Ba97,HC99,TWH99,TH99}; however this may be 
not the case at RHIC energies. 

Bose-Einstein (BE) correlation effects 
on the pion multiplicity distribution, on the single pion distribution, and on 
two-pion interferometry have been studied by many authors 
for a Gaussian source distribution\cite{multi,Z98,ZPH98,Zhang99,HPZ00}.  It has been shown 
by Bialas and Zalewski that those results are valid for a wide class 
of models\cite{BZ,Bialas,KZ}. Fialkowski and Wit have implemented multi-particle 
Bose-Einstein correlations in Monte-Carlo generators and 
have studied Bose-Einstein correlation effects on the $W$ mass shift, 
pion multiplicity distribution\cite{FW,FWW}. But 
the effects of 
multi-particle Bose-Einstein 
correlations on the mean phase space density 
have never been studied before. 
Theoretical studies\cite{HPZ00,Zhang99} have 
shown that pion interferometry depends strongly on 
the pion multiplicity distribution which was 
overlooked in previous HBT analyses. Thus it is 
interesting for us to study the effects of the 
general pion interferometry formula on the mean phase space density. 

This paper is 
arranged in the following way. In the Section II, 
 we re-derive Bertsch's formula and point out its implicit assumption. 
In Sec. III, we derive a phase space 
density  formula from the general pion interferometry and we 
find that if the phase space density becomes small this new expression 
 will be the same as the Bertsch's  except a extra normalization factor.
Unfortunately this simple relationship does not hold when the phase space 
density becomes large. In Sec. IV, multi-pion Bose-Einstein correlation 
effects on the mean phase space density are studied.  
We find that multi-particle BE correlations will increase the mean phase space density. 
Finally we give our conclusions 
in Sec. V.

\section{Bertsch's  formula and implicit assumptions}

The two-pion 
interferometry formula can be written as 
\begin{eqnarray}
C_2^{I}(p_1,p_2)&=&\frac{P_2(p_1,p_2)}{P_1(p_1)P_1(p_2)} =1+
\nonumber\\
&&\frac{\int d^4x d^4 y g^I(x,k)g^I(y,k)\exp(iq(x-y))}
{\int d^4x d^4 y g^I(x,p_1)g^I(x,p_2)}.
\label{e1}
\end{eqnarray}
Here $k=(p_1+p_2)/2$ and $q=p_1-p_2$ are two-pions average 
momentum and relative momentum respectively. $g^I(x,k)$ is a Wigner 
function which can be interpreted as the probability of finding a 
pion at position $x$ with momentum $k$. $P_2(p_1,p_2)$ and $P_1(p)$ 
are two-particle and single-particle inclusive distributions 
which are defined by 
\begin{eqnarray}
P_2(p_1,p_2)&=&\frac{d^6n}{d^3p_1d^3p_2}=
P_1(p_1)P_1(p_2)+
\nonumber\\
&&\int g^{I}(x,k)g^{I}(y,k)\exp(iq(x-y))d^4xd^4y, 
\nonumber\\
P_1(p)&=&\frac{d^3n}{d^3p}=\int d^4x g^{I}(x,p), 
\label{e2}
\end{eqnarray} 
with 
\begin{eqnarray}
\int d^3 p P_1(p)&=&\langle n\rangle,
\nonumber\\
\int d^3 p_1 d^3p_2  P_2(p_1,p_2)&=&\langle n(n-1)\rangle.
\label{e3}
\end{eqnarray}
From Eqs.~(\ref{e1},~\ref{e2}), it is easily checked that if\cite{Bertsch,TWH99}
\begin{equation}
g^{I}(x,k)\rightarrow \delta(x_0-\tau)f^{I}(x,k)\frac{1}{(2\pi)^3},
\label{e41}
\end{equation}
then 
\begin{eqnarray}
P_1(p)=\frac{1}{(2\pi)^3}\int f^{I}(x,p)d^3 x,
\label{e4}
\end{eqnarray}
and 
\begin{eqnarray}
P_2(p_1,p_2)-P_1(p_1)P_1(p_2)
&=&\frac{1}{(2\pi)^6}\int d^3x d^3 y f^{I}(x,k)
\nonumber\\
&& f^{I}(y,k) \exp(iq(x-y).
\label{e5}
\end{eqnarray}
The reason that we put $\frac{1}{(2\pi)^3}$ in Eq.~(\ref{e41}) is 
 that  in statistical physics for a infinite
volume\cite{ZP00} 
\begin{equation}
P_1(p)=\frac{V}{(2\pi)^3}f(p).
\label{e6}
\end{equation} 
Here $V$ is the volume and $f(p)$ is the Bose-Einstein 
distribution. 

Integrate Eq.(\ref{e5}) over $q$ we have 
\begin{eqnarray}
&&\int d^3 q [P_2(p_1,p_2)-P_1(p_1)P_1(p_2)]
\nonumber\\
&&=\frac{1}{(2\pi)^3}\int f^{I}(x,k)^2 d^3x.
\label{e7}
\end{eqnarray}

The average phase space density $\langle f^{I}\rangle_{k}$ 
can be  defined as 
\begin{eqnarray}
&&\langle f^{I}\rangle_k=\frac{\int f^{I}(x,k)^{2}d^3x}{\int f^{I}(x,k)d^3 x}
\nonumber\\
&&=\frac{\int d^3 q[P_2(k+\frac{q}{2},k-\frac{q}{2})-
P_1(k+\frac{q}{2})P_1(k-\frac{q}{2})]}
{P_1(k)}
\nonumber\\
&&
=\frac{\int d^3 q[C_2^I(q,k)-1]
P_1(k+\frac{q}{2})P_1(k-\frac{q}{2})}
{P_1(k)}
\label{e8}
\end{eqnarray}
Using the smooth approximation, $p_1\sim p_2\sim k$, which has been 
shown to be valid in heavy-ion collisions for its large 
phase space\cite{CGZ94}, we have 
\begin{equation}
\langle f^{I}\rangle_k=P_1(k)\int d^3 q[C_2^{I}(k,q)-1].
\label{e9}
\end{equation}
In Refs.\cite{Ba97,HC99,TWH99},
the authors have calculated the phase space density by assuming that  
\begin{eqnarray}
&&C_2^{I}(p_1,p_2)=1+
\nonumber\\
&&\lambda \exp(-\frac{1}{2}q_o^2R_o^2-
\frac{1}{2}q_s^2R_s^2-\frac{1}{2}q_l^2R_l^2-2R_oR_lq_oq_l).
\label{e11}
\end{eqnarray}

But the above parameterization of two-pion interferometry 
(Eqs.~(\ref{e11})) is not general, 
as in practice the two-pion 
correlation is fitted using function\cite{Zhang99,ZPH98}
\begin{equation}
C^{ex}_2(p_1,p_2)=AC_2^{I}(p_1,p_2).
\label{e15}
\end{equation}
Here $A$ is a normalization factor which exists in the two-pion 
interferometry formula\cite{Zhang99,ZPH98}. 
 If we use $C_2^{ex}(q,k)$ to take the place of $C_2^I(q,k)$ in 
Eq.~(\ref{e9}), the phase space density will be 
\begin{equation}
\langle f\rangle^{ex}_{k}=(A-1)P_1(k)+A\times \langle f^I\rangle_{k}.
\label{e17}
\end{equation}
Thus the phase space density will increase if 
$A$ is larger than one or decrease if $A$ is 
smaller than one.  In the latter  part of this  
paper we will show that this extra factor $A$ though 
has been used in the data analyses will not appear 
in the phase space density formula on the condition 
that the phase space is large and the density is 
small. This guarantee that the 
application of Bertsch formula for heavy-ion 
collisions system is appropriate if the phase space density is small.

\section{Phase space density from the general pion interferometry formula}

It  has been shown in 
Refs.~\cite{HPZ00,Zhang99} that 
the single particle spectrum, the two-particles spectrum and 
the two-pion interferometry formula read\cite{note1}
\begin{equation}
P_1(p)=\sum_{i=1}^{N_{max}}h_i G_i(p,p),
\label{e18}
\end{equation}
\begin{eqnarray}
P_2(p_1,p_2)&=&\sum_{i=1}^{N_{max}-1}\sum_{j=1}^{N_{max}-i}h_{i+j}
[G_i(p_1,p_1)G_i(p_2,p_2)
\nonumber\\
&&+G_i(p_1,p_2)G_i(p_2,p_1)].
\label{e19}
\end{eqnarray}
Where the definitions of $h_i$ and $G_i(p,q)$ can be found in Ref.\cite{HPZ00}. 
 $N_{max}$ is the maximum multiplicity 
 in the experiment. If $N_{max}=\infty$,  we will 
obtain the formula of $P_2(p_1,p_2)$ and 
$P_1(p)$ given in Refs.~\cite{HPZ00,Zhang99}. 

The two-pion correlation function is\cite{HPZ00}
\begin{eqnarray}
C_2(p_1,p_2)&=&\frac{P_2(p_1,p_2)}{P_1(p_1)P_1(p_2)}
\nonumber\\
&=&
C_2^{res}(p_1,p_2)[1+R_2(p_1,p_2)]
\label{e351}
\end{eqnarray}
with
\begin{eqnarray}
&&C_2^{res}(p_1,p_2)=
\nonumber\\
&&
\frac{\sum_{i}^{N_{max}-1}\sum_{j=1}^{N_{max}-i}
h_{i+j}G_i(p_1,p_1)G_j(p_2,p_2)}
{\sum_{i,j=1}^{N_{max}}h_ih_jG_i(p_1,p_1)G_j(p_2,p_2)},
\label{e34}
\end{eqnarray}
and
\begin{eqnarray}
&&R_2(p_1,p_2)=
\nonumber\\
&& \frac{\sum_{i=1}^{N_{max}-1}
\sum_{j=1}^{N_{max}-i}h_{i+j}G_i(p_1,p_2)G_j(p_2,p_1)}
{\sum_{i=1}^{N_{max}-1}\sum_{j=1}^{N_{max}-i}h_{i+j}G_i(p_1,p_1)G_j(p_2,p_2)}.
\label{e35}
\end{eqnarray}
In Ref.~\cite{HPZ00}, we have shown that $R_2(k,q)|_{q=\infty}=0$. 
So if the two-pion correlation function is 
expressed as
 Eq.~(\ref{e15}), then 
\begin{equation}
A=C_2^{res}(p_1,p_2),
\end{equation}
which is a function of $q$ and $k$.  We can always  
define a Wigner function $S(x,k)$ 
 which fulfils   
the following equation\cite{Z98,ZPH98,HPZ00}
\begin{equation}
\sum_{i=1}^{N_{max}}h_iG_i(p_1,p_2)=\int S(x,k)\exp(iqx) d^4x.
\label{e28}
\end{equation}
Thus,
\begin{equation}
P_1(p)=\sum_{i=1}^{N_{max}}h_iG_i(p,p)=\int S(x,p)d^4x.
\label{e40}
\end{equation}
It has been shown in 
Refs.~\cite{HPZ00,Z98,ZPH98} that for a special multiplicity 
distribution $p_n=\frac{\omega(n)}{\sum_{n=0}^{n=\infty} \omega(n)}$ 
or for a small phase space density and $p_n$ is a Poisson form, we have  
$h_{i+j}=h_ih_j$. Then Eqs.~(\ref{e19},~\ref{e351})  
change to (for $N_{max}=\infty$)\cite{Z98,ZPH98,HPZ00}
\begin{eqnarray}
P_2(p_1,p_2)&=&P_1(p_1)P_1(p_2)+
\nonumber\\
&&\int S(x,k)S(y,k)\exp(iq(x-y))dx dy,
\label{e411}
\end{eqnarray}
and 
\begin{eqnarray}
&&C_2(p_1,p_2)=1+
\nonumber\\
&&\frac{\int S(x,k)S(y,k)\exp(iq(x-y))dxdy}
{\int dx dy S(x,p_1)S(y,P_2)}.
\label{e42}
\end{eqnarray}
So we obtain Eq.~(\ref{e1}). Thus all the derivations given in Ref.\cite{Bertsch}
are valid. This implies that the normalization factor $A$ must be 
one and this can be shown from Eq. (\ref{e34}) under the 
condition that $N_{max}=\infty$ and $h_{i+j}=h_ih_j$.  
 From Eq.(\ref{e17}), we have 
\begin{equation}
\langle f\rangle^{ex}_{k}=\langle f^I\rangle_{k}.~~~
\end{equation}
This verifies the rightness of the application of Bertsch's formula 
for the case of small boson densities. 
However, the relationship $h_{i+j}=h_ih_j$ does not hold for all cases.
For the following four kinds of 
multiplicity distributions:
\begin{eqnarray}
p_n&=&\frac{\langle n\rangle^n}{n!}\exp(-\langle n\rangle)
\nonumber\\
&& ~~(Poisson~~~ distribution),
\nonumber\\
p_n&=&\frac{\langle n\rangle^n}{(1+\langle n\rangle)^{n+1}}~~~
\nonumber\\
&&~(Bose-Einstein~~~ distribution),
\nonumber\\
p_n&=&\frac{(n+k-1)!}{n!(k-1)!}
\frac{(\langle n\rangle/k)^n}{(1+\frac{\langle n\rangle}{k})^{n+k}} 
\nonumber\\
&&(negative~~ binomial~~ distribution),
\nonumber\\
p_n&=&\frac{1}{n\Gamma(k)}(\frac{kn}{\langle n\rangle})^{k}\exp(-kn/\langle n\rangle)~~
\nonumber\\
&&~~(Gamma~~ distribution),
\end{eqnarray}
We can 
prove that\cite{HPZ00} that 
if the phase space volume is large, 
$h_2/h_1^2\sim h_3/(h_1h_2)\sim 1$ for the Poisson distribution, 
$h_2/h_1^2=2$ and $h_3/(h_1h_2)=3$ for the Bose-Einstein 
distribution. 


If there are strong BE correlations among bosons, then  it is impossible to 
express  
the two-particles distribution $S(x,p_1;y,p_2)$ as 
$S(x,p_1)S(y,p_2)$. However,  
one can find a real function $S_i(x,k)$ which fulfils  
the following equation
\begin{equation}
G_i(p_1,p_2)=\int S_i(x,k)\exp(iqx)d^4x.
\label{e33}
\end{equation}
Then we define $S(x,k;y,k)$ 
and $S(x,p_1;y,p_2)$ as  
\begin{eqnarray}
S(x,k;y,k)&=&\sum_{i=1}^{N_{max}-1}\sum_{j=1}^{N_{max}-i}h_{i+j}S_i(x,k)S_j(y,k),
\label{e37}
\end{eqnarray}
and
\begin{eqnarray}
S(x,p_1;y,p_2)&=&\sum_{i=1}^{N_{max}-1}\sum_{j=1}^{N_{max}-i}
\nonumber\\
&&
h_{i+j}S_i(x,p_1)S_j(y,p_2),
\label{e38}
\end{eqnarray}
which satisfies
\begin{eqnarray}
&&\sum_{i=1}^{N_{max}-1}\sum_{j=1}^{N_{max}-i}h_{i+j}G_i(p_1,p_2)G_j(p_2,p_1)
=
\nonumber\\
&&\int S(x,k;y,k)\exp(iq(x-y))dxdy,
\label{e36}
\end{eqnarray}
and 
\begin{eqnarray}
&&\sum_{i=1}^{N_{max}-1}\sum_{j=1}^{N_{max}-i}
h_{i+j}G_i(p_1,p_1)G_j(p_2,p_2)
=
\nonumber\\
&&\int S(x,p_1;y,p_2)dxdy.
\end{eqnarray}
Because $\rho^{*}(p_1,p_2)=\rho(p_2,p_1)$, then 
$G_i^{*}(p,p)=G_i(p,p)$ and $G_i^{*}(p,q)=G_i(q,p)$. Thus  
$S(x,p_1;y,p_2)$ must be a real function which fulfils the 
requirement of the Wigner function. If 
$h_{i+j}=h_ih_j$ and $N_{max}\rightarrow \infty$,
from Eq.~(\ref{e38}), Eq.~(\ref{e33}) and 
Eq.~(\ref{e28}), we have $S(x,p_1;y,p_2)=S(x,p_1)S(y,p_2)$, 
thus Eqs.~(\ref{e40},\ref{e411},\ref{e42}) are obtained. So we can identify 
$S(x,p_1,y,p_2)$ as a two pion distribution function which represents the 
probability of  two pions  
emitted from point $x$ with momentum $p_1$ and from point $y$ 
with momentum $p_2$ respectively.  Two-particle spectrum 
distributions can be written as
\begin{eqnarray}
P_2(p_1,p_2)&=&\int S(x,p_1;y,p_2)dx dy +
\nonumber\\
&&
\int S(x,k;y,k)\exp{[iq(x-y)]}dxdy.
\end{eqnarray}
Then $R_2(p_1,p_2)$ reads
\begin{eqnarray}
R_2(p_1,p_2)=\frac{\int S(x,k;y,k)\exp(iq(x-y))dx dy}
{\int S(x,p_1;y,p_2) dx dy}.
\end{eqnarray} 
If\cite{Bertsch}
\begin{eqnarray}
S(x,k;y,k)&\rightarrow &
\delta(x_0-\tau)\delta(y_0-\tau)f(x,k;y,k)\frac{1}{(2\pi)^6}
\label{e41x}
\end{eqnarray}
and 
\begin{eqnarray}
S(x,k)&\rightarrow &\delta(x_0-\tau)\frac{1}{(2\pi)^3}f(x,k),
\label{e4211}
\end{eqnarray}
then 
\begin{eqnarray}
\langle f\rangle_{k}&=&
\frac{\int f(x,k;x,k) d^3x}{\int f(x,k) d^3x}
\nonumber\\
&=&\frac{\int d^3q R_2(p_1,p_2)C_2^{res}(p_1,p_2)P_1(p_1)P_1(p_2)}
{P_1(k)}.
\label{f40}
\end{eqnarray}
Using the smoothness approximation, 
$P_1(p_1)\sim P_1(p_2)\sim P_1(k)$, we have 
\begin{eqnarray}
\langle f\rangle_{k}&=&P_1(k)\int d^3q [\frac{C_2(p_1,p_2)}{C_2^{res}(p_1,p_2)}-1]
C_2^{res}(p_1,p_2)
\nonumber\\
&=&AP_1(k)\int d^3q[\frac{C_2(q,k)}{A}-1]
\nonumber\\
&=&A\langle f^{I}\rangle_{k}.
\label{e49}
\end{eqnarray}
In the above we have used the approximation 
$A\sim C_2^{res}(q,k)\sim const$. 
It is interesting 
to point out that normally $A$ should be a function of $q$ and $k$; 
however, in heavy-ion collisions, the practice is to 
normally fit it as a constant if the phase space density 
is small\cite{HPZ00}.
 If the phase space density is high, it has been suggested in Ref.\cite{HPZ00} 
to fit data using the function
\begin{eqnarray}
C_2(p_1,p_2)&=&C_2^{res}(p_1,p_2)[1+R_2(p_1,p_2)].
\label{e51}
\end{eqnarray}
Here  
$C_2^{res}(q,k)={\cal{N}}[1+B(k)\cdot exp(-q^2 R_{res}^2(k))$ and 
$R_2=\lambda(k) \exp(-q^2 R^2(k))$. 
Thus it is not the best choice to use a constant $A$ 
in Eq.(\ref{e15}).

\section{multi-particle BE correlation effects on the mean phase space 
density}
In the following we will study the effects of 
multi-pion 
BE correlations  on the mean phase space density. 
We assume   $g^I(x,p)$ to be\cite{multi,PGG90} 
\begin{equation}
g^I(x,p)=\delta(x_0)\frac{n_0}{(2\pi R\Delta )^3}
\exp(-\frac{x^2}{2R^2}-\frac{p^2}{2\Delta^2}),
\label{e31}
\end{equation}
then $f^{I}(x,p)$ reads
\begin{equation}
f^{I}(x,p)=\frac{(2\pi)^3n_0}{(2\pi R\Delta)^3}\exp(-\frac{x^2}{2R^2}-\frac{p^2}{2\Delta^2}).
\label{e32}
\end{equation}
Due to Eq.~(5), we immediately come to the conclusion that 
$n_0$ is the mean pion multiplicity observed in the experiment.
It is easily checked that 
\begin{equation}
\langle f^{I}\rangle_{k}=\frac{\int d^3x f^{I}(x,p)^2}{\int d^3x f^{I}(x,p)}=
\frac{n_0}{(\sqrt{2}R\Delta)^3}\exp(-\frac{p^2}{2\Delta^2}).
\label{e32y}
\end{equation}
However, we have  
neglected the high-order BE correlation effects to get this 
$\langle f^I \rangle_k$. If we 
keep only the leading terms in Eqs. (19,20)(correspondingly, this 
implies that we have assumed that the phase space volume is 
 large), then 
\begin{equation}
P_1(p)=h_1G_1(p,p),
\end{equation}
and 
\begin{eqnarray}
P_2(p_1,p_2)&=&h_2[G_1(p_1,p_1)G_1(p_2,p_2)+
\nonumber\\
&&G_1(p_1,p_2)G_1(p_2,p_1)].
\end{eqnarray}
Thus the corresponding two-pion correlation function becomes 
\begin{equation}
C_2(p_1,p_2)=\frac{h_2}{h_1^2}[1+\frac{G_1(p_1,p_2)G_1(p_2,p_1)}{
G_1(p_1,p_1)G_1(p_2,p_2)}]  .
\label{ext51}
\end{equation}
Bring Eq.~(\ref{ext51}) into Eq.~(\ref{e8}) and ignore the 
extra normalization 
factor, $\frac{h_2}{h_1^2}$, we get Eq.~(\ref{e32y}); furthermore, 
if we bring 
Eq.~(\ref{ext51}) into  
Eq.~(\ref{f40}) and ignore the extra normalization factor,
 Eq.~(\ref{e32y}) will be 
regained. This results demonstrate that the Bertsch formula 
 will give the correct result on the condition that 
the phase space volume is large.  In the following we will 
study the multi-particles correlation effects on the 
mean phase space density due to the large phase space density. 
 
\begin{figure}[h]\epsfxsize=8cm
\centerline{\epsfbox{fig.1a}}
\end{figure} 
\begin{figure}[h]\epsfxsize=8cm
\centerline{\epsfbox{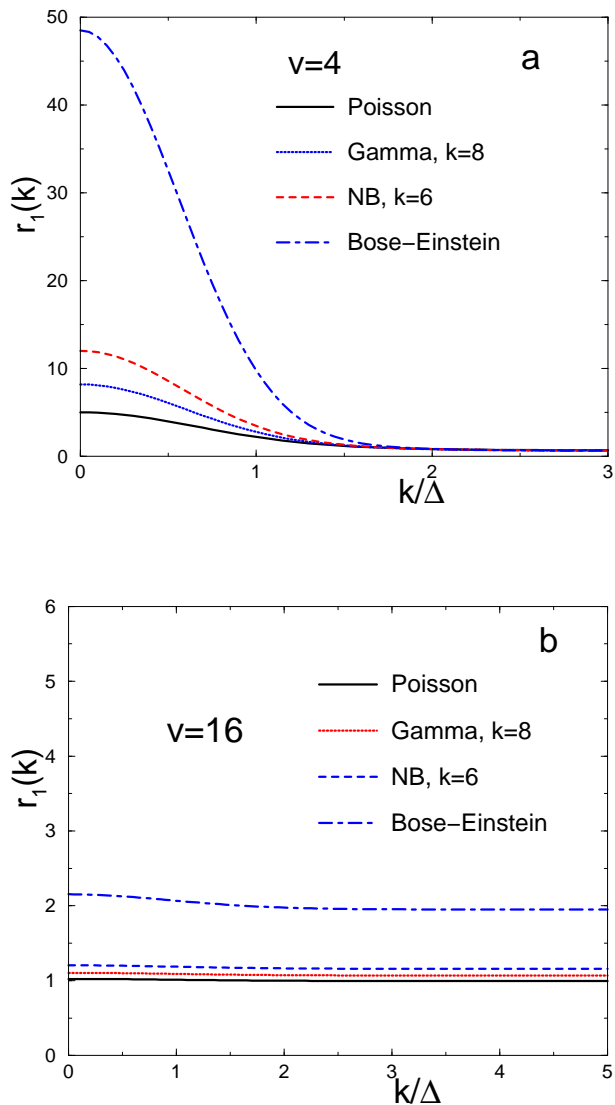}}
\caption{\it $r_1(k)$ as a function of $k/\Delta$. 
 Here the mean pion multiplicity is $20$.
The solid line, dashed line, dotted line and 
dot-dashed lines correspond to Poisson, 
Gamma, Negative binomial, and Bose-Einstein distribution.
The phase space $v=2R\Delta=4$ and $16$ respectively.  }
\label{f5}
\end{figure} 

We define a function $r_1(k)$ as
\begin{equation}
r_1(k)=\frac{\langle f\rangle_{k}}{\langle f^{I}\rangle_{k}}.
\end{equation}

In the Fig.\ref{f5}, $r_1(k)$  
is shown as a function of $k/\Delta$. 
  It is clear that 
for a large phase space,  
$r_1(k) \sim constant$;  on the other 
hand, if the phase space is small, $\langle f\rangle_{k}$ 
and $\langle f^{I}\rangle_{k}$ 
have big differences at small momentum and small 
differences at large momentum. This is easily understood 
since quantum effects are  big for small momentum particles.
It is interesting to notice that when the phase space is 
large, $r_1\sim h_2/h_1^2(\sim C^{res}(q,k))$, which is 
two for the Bose-Einstein distribution,
one for the Poisson distribution.
From Eq.~(\ref{e41x}) and Eq.~(\ref{e36})(taking $q=0$), we have 
\begin{eqnarray} 
&&\int f(x,k;y,k)d^3x d^3y=(2\pi)^6\int S(x,k;y,k)d^4x d^4y
\nonumber\\
&=&(2\pi)^6\sum_{i=1}^{N_{max}-1}\sum_{j=1}^{N_{max}-i}
h_{i+j}G_i(k,k)G_j(k,k).
\label{e51z}
\end{eqnarray} 
From Eq.~(\ref{e4211}) and Eq.~(\ref{e28})(taking $q=0$), we get 
\begin{eqnarray}
\int f(x,k)dx&=&(2\pi)^3\int S(x,k) d^4x
\nonumber\\
&=&(2\pi)^3\sum_{i=1}^{N_{max}}h_iG_i(k,k).
\label{e51y}
\end{eqnarray}
From the definition 
of $C_2^{res}(p_1,p_2)$(Eq.~(\ref{e34})) and 
Eqs.~(\ref{e51z},~\ref{e51y}), we have 
\begin{equation}
C_2^{res}(q,k)_{q=0}=\frac{\int f(x,k;y,k) d^3x d^3y}{\int d^3x d^3y f(x,k)f(y,k)}.
\end{equation}

\begin{figure}[h]\epsfxsize=8cm
\centerline{\epsfbox{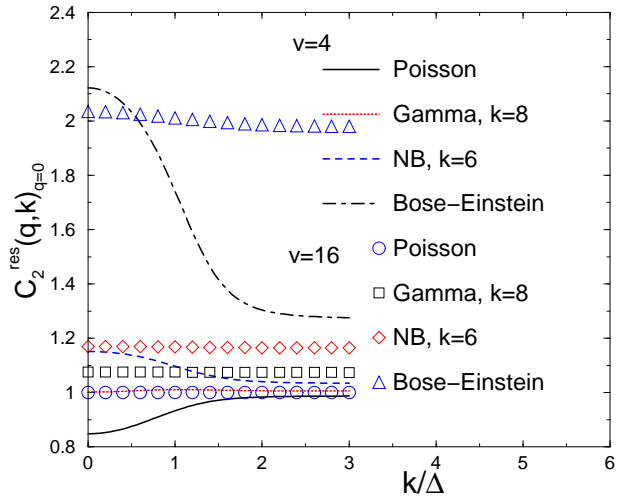}}
\caption{\it $C_2^{res}(q,k)|_{q=0}$ vs. $k/\Delta$. 
 Here the mean pion multiplicity is $20$.
The solid line (circles), dashed line (squares), dotted line
(diamonds) and 
dot-dashed lines (triangles) corresponds to Poisson, 
Gamma, Negative binomial, and Bose-Einstein distribution.
The phase space $v=2R\Delta=4$ and $16$ respectively.  }
\label{f41}
\end{figure} 

In Fig.~\ref{f41}, $C_2^{res}(q,k)_{q=0}$ vs 
$k/\Delta $ is shown. 
 We notice that when phase space is 
large, $C_2^{res}(q,k)_{q=0}$ is 
a constant. In this case, 
$\int d^3x d^3y f(x,k;y,k) = 2 \int d^3 d^3 y f(x,k) f(y, k)$ for 
the Bose-Einstein distribution and 
$\int d^3x d^3y f(x,k;y,k) = \int d^3 d^3 y f(x,k) f(y, k)$ for
the Poisson multiplicity distribution. But these relationships 
do not hold anymore for a small phase space volume. 

In Eqs.~(\ref{e8},\ref{e49}), we find that $\langle f\rangle_{k}$ 
 is $A$ times 
larger than $\langle f^{I}\rangle_{k}$ when the 
phase space density is small and the function form of 
$\langle f\rangle_k$ and $\langle f^I\rangle_k$ are different 
in the numerator. In the following we 
would like to show the relationship between  
$f(x,k;x,k)$ and $f(x,k)$. It is found that 
\begin{eqnarray}
&&\int f(x,k;x,k)d^3x=
\nonumber\\
&&\frac{1}{(2\pi)^3}\int f(x,k;y,k)e^{i{\bf q}({\bf x-y})}d^3q d^3xd^3y
\nonumber\\
&=&(2\pi)^3\int S(x,k;y,k)e^{iq(x-y)} d^4x d^4y d^3q
\nonumber\\
&=&(2\pi)^3\int d^3q 
\sum_{i=1}^{N_{max}-1}
\sum_{j=1}^{N_{max}-i}
\nonumber\\
&&h_{i+j}G_i(p_1,p_2)G_j(p_2,p_1)
\end{eqnarray}
and 
\begin{eqnarray}
\int f^2(x,k)d^3x &=&(2\pi)^3
\int d^3q 
\sum_{i=1}^{N_{max}}
\sum_{j=1}^{N_{max}}
\nonumber\\
&&
h_{i}h_{j}G_i(p_1,p_2)G_j(p_2,p_1).
\end{eqnarray}
In the above derivation, we have used 
Eqs.~(\ref{e28},~\ref{e33},~\ref{e36},~\ref{e41x},~\ref{e4211}). 
We define a function $r_2(k)$ as 
\begin{equation}
r_2(k)= \frac{\int f(x,k;x,k)d^3x }{\int d^3x f^2(x,k)}.
\end{equation}
Similar to Ref.\cite{HPZ00}, we can prove that 
\begin{equation} 
r_2(k)\sim C_2^{res}(q,k)\sim \frac{h_2}{h_1^2}=constant~~~~v\rightarrow\infty~ .  
\label{e51x}
\end{equation}

\begin{figure}[h]\epsfxsize=8cm
\centerline{\epsfbox{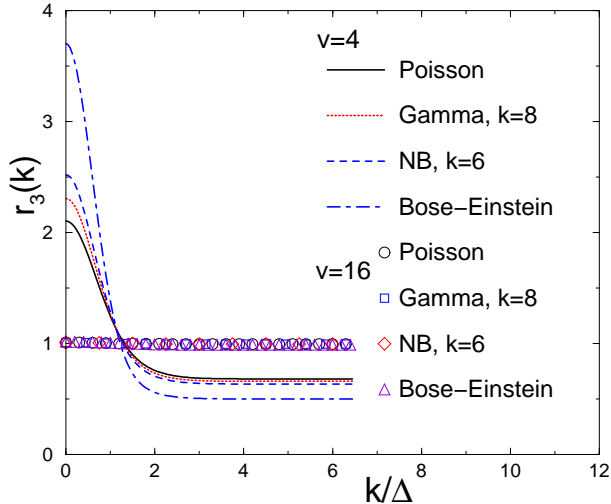}}
\caption{\it $r_3(k)$ as a function of $k/\Delta$. 
 Here the mean pion multiplicity is $20$.
The solid line (circles), dashed line (squares), dotted line
(diamonds) and 
dot-dashed lines (triangles) corresponds to Poisson, 
Gamma, Negative binomial, and Bose-Einstein distribution.
The phase space $v=2R\Delta=4$ and $16$ respectively.  }
\label{f7}
\end{figure}

In the following,  we will discuss the effects of multi-boson 
correlations on the distribution function of pions in 
the momentum space.  If the multi-boson 
symmetrization effects are small, the distribution function is 
$f^{I}(x,p)$. If the  multi-boson correlations are strong, the  
distribution function is $f(x,k)$. We define 
a function 
\begin{equation}
r_3(k)=\frac{\int d^3x f(x,k)}{\int f^{I}(x,k)d^3x}
=\frac{(2\pi)^3 P_1(k)}{\int f^{I}(x,k) d^3x},
\end{equation}
which reflects the effects of multi-boson correlations on the 
distribution function.  
In Fig.\ref{f7}, $r_3(k)$ vs. $k/\Delta$ is shown. 
It is interesting to notice that when phase space is 
large, 
\begin{equation}
r_3(k)=1, v\rightarrow \infty
\label{e56}
\end{equation}
for all distributions.  On the other hand, when phase space is small, 
boson density becomes large at small momentum region. 
  We define 
$\langle f\rangle^I_{k}$ as 
\begin{equation}
\langle f\rangle^{I}_{k}=\frac{\int f^2(x,k)dx }{\int f(x,k)dx}.
\end{equation}
This definition is similar to Eq.~(\ref{e8}) but with $f(x,k)$ taking 
the 
place of $f^I(x,k)$.  
From Eq.~(\ref{e8}), Eq.~(\ref{e15}), Eq.~(\ref{f40}), 
 Eq.~(\ref{e49}), and Eq.~(\ref{e51x}), 
we get 
\begin{eqnarray}
\langle f\rangle^{I}_{k}&= & \frac{\langle f\rangle_k}{r_2(k)}
\nonumber\\
&\sim &P(k)\int d^3 q[\frac{C_2(q,K)}{A}-1]~~v\rightarrow \infty.
\label{e52}
\end{eqnarray}
This is one of the main results of this paper. This result guarantees that 
the application of Bertsch's formula for a large system and 
dilute gas is appropriate. 
On the other hand, Bertsch formula is incomplete when the phase space 
density is high. This comes from the fact that we can not calculate 
$\int f^2(x,k) d^3x$ from two-pion interferometry formula though we 
could calculate $\int f(x,k)d^3x$ from the single particle spectrum. 
If the phase space density is high,  
 $C_2^{res}(p_1,p_2)$ is not a constant anymore, we can not 
find the approximation formula as Eq.~(\ref{e52}); however 
 Eq.~(\ref{f40}) can still be used to find the ratio of  
 the number of particles pairs which are emitted from the same phase space cell 
 to the average number of particles. The relation among  
 $\langle f\rangle_{k}$, $\langle f\rangle^I_k$ and $\langle f^I\rangle_k$ 
 will be very complex when the phase space density is large. 
Then the physical meaning of the Bertsch formula is no longer clear.
Multi-pion BE symmetrization 
will affect the current formalism and we believe that those 
effects will be similar to the effects of multi-pion BE on 
the Wigner function $g(x,p)$, which have been presented in Ref.\cite{ZPH98}
(or $f^I(x,p)$ presented here). 

\section{Conclusions}

In this paper,  the mean phase space density distribution 
of  bosons is derived from the general pion 
interferometry formula. We find that when the phase space is small 
and thus the 
boson density is high, the two particle source 
distribution can not be expressed as a product of two single particles source distributions.
On the other hand,
when the phase space is large and thus the boson density is small, Bertsch's  formula 
is  recovered. Thus Bertsch's formula 
can be used for the heavy-ion system if the freezeout pion 
phase space density is small. Multi-pion BE correlation effects on the mean 
phase 
space density distribution are studied, it is found that when 
the phase space density is large, bosons are concentrated in  
small momentum region and this effect is connected 
with the pion multiplicity distributions.

\begin{center}
{\bf Acknowledgement}
\end{center}

This work was supported in part by the 
Natural Science and Engineering Research Council of Canada, and part by the
Fonds Nature et Technologies of Quebec.


\begin{thebibliography}{29}
\bibitem{note1}
In previous writting, one normally sums the terms 
in Eqs.(\ref{e18},\ref{e19}) from one 
to $\infty$. But, in pratice, one can change the summation 
from $N_{min}$ to $N_{max}$.  In this paper, we choose  
 $N_{min}=1$ for simplicity and this does not affect our conclusions.
\end{thebibliography}
\end {document}